\documentstyle[12pt]{article}
\begin{document}

QUANTUM BIRTH OF A HOT UNIVERSE\\
\bigskip

I.G. Dymnikova\\
Institute of Mathematics, Informatics and Physics, University of Olsztyn,\\
\.Zo{\l}nierska 14, 10-561, Olsztyn, Poland;\\
e-mail: irina@matman.uwn.edu.pl\\

M.L. Fil'chenkov\\
Alexander Friedmann Laboratory for Theoretical Physics,\\
St. Petersburg, Russia\\
Institute of Gravitation and Cosmology, Peoples' Friendship University\\
of Russia, 6 Miklukho-Maklaya Street, Moscow 117198, Russia;\\
e-mail: fil@agmar.ru\\

\begin{abstract}

We consider quantum birth of a hot Universe in the framework of quantum
geometrodynamics in the minisuperspace model. The energy spectrum of the
Universe in the pre-de-Sitter domain naturally explains the cosmic microwave
background (CMB) anisotropy. The false vacuum where the Universe tunnels from
the pre-de-Sitter  domain is assumed to be of a Grand Unification Theory
(GUT) scale. The probability of the birth of a hot Universe from a quantum
level proves to be about $10^{-10^{14}}$. In the presence of matter with a
negative pressure (quintessence) it is possible for open and flat universes
to be born as well as closed ones.
\end{abstract}

\section{Introduction}

In the framework of the standard scenario\quad\cite{DZS,L} a quantum birth of
the Universe\quad\cite{T,F}, as a result of tunnelling\quad\cite{V}, is
followed by a classical decay of the de Sitter (false) vacuum with the
equation of state $p=-\varepsilon$ into a hot expanding Universe called the
Big Bang. One of the proofs of the hot Universe model is a discovery of the
CMB with the temperature about $3 K$\quad\cite{P}. G. Gamow, the author of
the tunnel effect\quad\cite{GZP} basic to the Universe tunnelling, was the
first to predict CMB\quad\cite{GPR} being a radiation of the hot Universe
cooled due to its expansion. Recently a CMB anisotropy $\frac{\Delta T}{T}
\simeq 10^{-5}$ has been discovered\quad\cite{S}. This anisotropy allows a
large-scale structure to be explained.

In the present paper we consider a quantum model of the hot Universe. In the
pre-de-Sitter domain radiation energy levels are quantized, which allows
temperature fluctuations to be treated as a manifestation of a quantum
behaviour of the Universe before its birth from the false vacuum. In the
previous papers\quad\cite{F5,F8} as well as in\quad\cite{VG,Kz} the presence
of a nonzero energy in Schr\"odinger's equation was shown to be due to
radiation or ultrarelativistic gas. Here we shall calculate the temperature
of this radiation as well as the probability of its tunnelling through the
barrier separating the pre-de-Sitter domain from the false vacuum. The
quantized temperature is compared with the observed CMB anisotropy. We assume
that the false vacuum energy density is at the GUT scale.

\section{Approach}

The Wheeler-DeWitt equation for Friedmann's world reads\quad\cite{F5}
$$
\frac{d^2\psi}{da^2}-V(a)\psi=0                          \eqno(1)
$$
where
$$
V(a)=\frac{1}{l_{pl}^4}\left(ka^2-\frac{8\pi G\varepsilon a^4}{3c^4}\right),
                                                                     \eqno(2)
$$
$a$ is the scale factor, $k=0,\pm 1$ is the model parameter.\\
As follows from Einstein's equations, the energy density may be written
in the form
$$
\varepsilon=\varepsilon_0\sum_{n=0}^6 B_n\left(\frac{r_0}{a}\right)^n.\eqno(3)
$$
Here $B_n$ are contributions of different kinds of matter to the total
energy density at the de Sitter horizon scale.
$$
n=3(1+\alpha)                                                   \eqno(4)
$$
where $\alpha$ is the parameter characterizing the equation of state
$$
p=\alpha\varepsilon.                                              \eqno(5)
$$
For the equations of state satisfying the weak energy dominance condition
$$
|p|\le\varepsilon,\quad \varepsilon>0                              \eqno(6)
$$
we have:\\
$n=0\quad (\alpha=-1)$ for the de Sitter (false) vacuum,\\
$n=1\quad (\alpha=-\frac{2}{3})$ for domain walls,\\
$n=2\quad (\alpha=-\frac{1}{3})$ for strings,\\
$n=3\quad (\alpha=0)$ for dust,\\
$n=4\quad (\alpha=\frac{1}{3})$ for radiation or ultrarelativistic gas,\\
$n=5\quad (\alpha=\frac{2}{3})$ for perfect gas,\\
$n=6\quad (\alpha=1)$ for ultrastiff matter.\\
Matter with a negative pressure (in particular, corresponding to the equations
of state for false vacuum, domain walls and strings) has been recently called
quintessence \quad\cite{CDS} due to which the Universe expands with
acceleration.

The de Sitter horizon is defined as
$$
\frac{1}{r_0^2}=\frac{8\pi G\varepsilon_0}{3c^4}.                  \eqno(7)
$$
Since $\varepsilon=\varepsilon_0$ at $a=r_0$, we obtain
$$
\sum_{n=0}^6 B_n=1.                                                 \eqno(8)
$$
Separating in the potential (2) a term independent of the scale factor, we
reduce the Wheeler-DeWitt equation to Schr\"odinger's
$$
-\frac{\hbar^2}{2m_{pl}}\frac{d^2\psi}{da^2}-[U(a)-E]\psi=0         \eqno(9)
$$
for a planckeon with the energy $E$, corresponding to radiation, moving in
the potential created by other kinds of matter. The potentials in
Wheeler-DeWitt's and Schr\"odinger's equations are related by the formula
$$
V(a)=\frac{2m_{pl}}{\hbar^2}[U(a)-E].                               \eqno(10)
$$
Restricting ourselves to radiation, strings and the de Sitter vacuum, we
obtain
$$
U(a)=\frac{m_{pl}c^2}{2l_{pl}^2}[(k-B_2)a^2-\frac{B_0a^4}{r_0^2}], \eqno(11)
$$
$$
E=\frac{m_{pl}c^2}{2}\left(\frac{r_0}{l_{pl}}\right)^2B_4.         \eqno(12)
$$
The energy $E$ is related to the contribution of radiation to the total
energy density.

\section{WKB Calculation of the Energy Spectrum and Penetration Factor}

The quantization of energy in the well (a Lorentzian domain of the
pre-de-Sitter Universe) follows the Bohr-Sommerfeld formula\quad\cite{LQM}
$$
2\int\limits_0^{a_1}\sqrt{2m_{pl}(E-U)}\,da=\pi\hbar \left(n+\frac{1}{2}
\right),                                                            \eqno(13)
$$
where $U(a_1)=E,\quad n=1, 3, 5$,...(since $\psi(0)=0$ if $U=\infty$ for
$a<0)$, and the penetration factor for the Universe tunnelling through the
potential barrier between the pre-de-Sitter and de Sitter domains is given by
Gamow's formula
$$
D=\exp\left(-\frac{2}{\hbar}|\int\limits_{a_1}^{a_2}\sqrt{2m_{pl}
(E-U)}\,da|\right)                                                \eqno(14)
$$
where $U(a_1)=U(a_2)=E$.

Mathematically, the problem reduces to evaluation of $\int\limits_{}^{}\sqrt
{2m_{pl}(E-U)}\,da$ where the energy and the potential satisfy formulae (11),
(12) respectively. The potential (11) has a minimum $U=0$ at $a=0$, a maximum
$U=\frac{m_{pl}c^2}{8B_0}(k-B_2)^2\left(\frac{r_0}{l_{pl}}\right)^2$ at $a=r_0
\sqrt{\frac{k-B_2}{2B_0}}$ and zeros at $a=0$ and $r_0=\sqrt{\frac{k-B_2}{B_0}
}$, where $k-B_2>0$ and $B_0>0$.

Near the minimum we have
$$
U=U(0)+\frac{1}{2}\frac{d^2U}{da^2}|_{a=0}\cdot a^2                \eqno(15)
$$
for $a\ll r_0\sqrt{\frac{k-B_2}{B_0}}$.

Near the maximum we have
$$
U=U(a_{max})+\frac{d^2U}{da^2}|_{a=a_{max}}\cdot(a-a_{max})^2       \eqno(16)
$$
for $|r_0\sqrt{\frac{k-B_2}{2B_0}}-a|\ll r_0\sqrt{\frac{k-B_2}{2B_0}}$.

Formulae (13) and (14) take the same value
$$
U_{med}=\frac{m_{pl}c^2}{2}(k-B_2)^2\left(\frac{r_0}{l_{pl}}\right)^2[\frac
{1}{(1+\sqrt{2})^2}-\frac{1}{(1+\sqrt{2})^4}]                      \eqno(17)
$$
at
$$
a_{med}=\frac{r_0}{1+\sqrt{2}}\sqrt{\frac{k-B_2}{B_0}}.             \eqno(18)
$$
Hence $U_{med}=4[\frac{1}{(1+\sqrt{2})^2}-\frac{1}{(1+\sqrt{2})^4}]U_{max}
\approx 0.569 U_{max}$ at $a_{med}=\frac{\sqrt{2}}{1+\sqrt{2}}a_{max}
\approx 0.586a_{max}$. Thus we may use formula (15) for $a\le 0.586 a_{max}$
and $U\le 0.569 U_{max}$ and formula (16) for $a\ge 0.586 a_{max}$ and
$U\ge 0.569 U_{max}$.

Using formulae (13) and (15), we calculate the energy spectrum
$$
E=m_{pl}c^2\sqrt{k-B_2}\left(n+\frac{1}{2}\right)                \eqno(19)
$$
where
$$
n+\frac{1}{2}<\frac{(k-B_2)^{3/2}}{8B_0}\left(\frac{r_0}{l_{pl}}\right)^2
                                                                  \eqno(20)
$$
since $E<U_{max}$. Although formula (19) has been obtained in the WKB
approximation, it coincides with the exact solution for a harmonic oscillator
considered previously for the case $r_0=l_{pl}$\quad\cite{F5}.

Using formulae (14) and (16), we calculate the penetration factor near the
maximum of the potential
$$
D=\exp\left\{-\pi\left(\frac{r_0}{l_{pl}}\right)^2\frac{|\frac{(k-B_2)^2}
{4B_0}-B_4|}{\sqrt{2(k-B_2)}}\right\}.                            \eqno(21)
$$
Although the problem of penetration through a barrier near its maximum was
considered by other authors\quad\cite{LQM,KM}, our approach gives a more
exact formula because we do not expand $\sqrt{2m_{pl}(E-U)}$ in series for a
parabolic potential and then calculate $\int\limits_{}^{}\sqrt{2m_{pl}(E-U)}\,
da$ but calculate this integral directly.

For $B_4\ll \frac{(k-B_2)^2}{4B_0}$ the penetration factor (21) reduces to
$$
D=\exp\left\{-\frac{2}{3}\frac{(k-B_2)^{3/2}}{B_0}\left(\frac{r_0}{l_{pl}}
\right)^2\right\}.                                                \eqno(22)
$$

Formulae (21) and (22) satisfy the WKB approximation as $\left(\frac{r_0}{l_
{pl}}\right)^2\gg 1$. As seen from them, open and flat universes can be born
if $k-B_2>0$, i.e for $B_2<0$, in other words for quintessence with a
negative energy density.

\section{Cosmic Microwave Background Temperature and Anisotropy. Probability
of the Birth of a Hot Universe}

In the hot Universe model the energy density of radiation and
ultrarelativistic gas is given by the formula\quad\cite{LFT}
$$
\varepsilon=\frac{3c^2}{32\pi Gt^2}.                             \eqno(23)
$$
On the other hand for the matter with the equation of state $p=\frac
{\varepsilon}{3}$ we have\quad\cite{L}
$$
\varepsilon=\frac{4}{c}\sigma\Theta^4N(\Theta)                    \eqno(24)
$$
where $\sigma=\frac{\pi^2}{60\hbar^3c^2}$,\quad $\Theta$ is the temperature
in degrees $T$ multiplied by the Bolzmann constant (the average energy of a
particle $\bar E=3\Theta$), $N(\Theta)=10^2 - 10^4$ is assumed to be
determined from observations. From formulae (23), (24) we obtain\quad\cite{ZN}
$$
\Theta=\sqrt[4]{\frac{45}{32\pi^3N(\Theta)}}m_{pl}c^2\sqrt{\frac{t_{pl}}{t}}
                                                                \eqno(25)
$$
where $N(\Theta)=4.07\cdot 10^3$, which gives $T=2.73$ K for $t=1.5\cdot 10^
{10}$yr (the Hubble constant $H_0=65$ km$\cdot$ s${}^{-1}\cdot$ Mps${}^{-1}$).

Assume that the false vacuum energy density is related to Grand Unification
scale $E_{GUT}$ which can be described by the formula\quad\cite{R}
$$
E_{GUT}=m_pc^2e^{\frac{\hbar c}{4e^2}}=7.03\cdot 10^{14} \rm GeV     \eqno(26)
$$
where $m_p$ is the proton mass. Substituting (26) into (25), and taking
account $3\Theta(t_0)=E_{GUT}$, we obtain $t_0=4.86\cdot 10^{-37}$ s and
$r_0=2.92\cdot 10^{-26}$ cm.

The energy (19) is the energy density (24) multiplied by the volume
$\frac{4\pi}{3}r_0^3$. It gives us the quiantized temperature
$$
\Theta=\sqrt[4]{\frac{45}{4\pi^3N(\Theta)}}\sqrt[8]{k-B_2}\left(\frac{l_{pl}}
{r_0}\right)^{3/4}m_{pl}c^2\left(n+\frac{1}{2}\right)^{1/4}.  \eqno(27)
$$
The average energy of a particle is estimated within the range
$$
1.42\cdot 10^{13}\rm GeV\le \bar E\le 3.24\cdot 10^{16}\rm GeV
$$
for
$$
\frac{3}{2}\le n+\frac{1}{2}\le\frac{(k-B_2)^{3/2}}{8B_0}\left(\frac{r_0}
{l_{pl}}\right)^2.
$$
The lowest energy is close to the values predicted by reheating models, the
highest one, being the most probable, is of the order of the monopole rest
energy\quad\cite{L}. Thus the model predicts existence of monopoles at the
beginning of inflation which dilutes their density to the required level.

On the other hand, equating (12) to (19), we obtain
$$
B_4=2\sqrt{k-B_2}\left(\frac{l_{pl}}{r_0}\right)^2\left(n+\frac{1}{2}\right).
                                                        \eqno(28)
$$
CMB temperature fluctuations are given by the formula
$$
\frac{\Delta T}{T}=\frac{\sqrt[4]{n+\frac{3}{2}}-\sqrt[4]{n+\frac{1}{2}}}
{\sqrt[4]{n+\frac{1}{2}}}.                                  \eqno(29)
$$
For $n\gg 1$ we have
$$
\frac{\Delta T}{T}\approx\frac{1}{4n}.                          \eqno(30)
$$
For $n=2.5\cdot 10^4$ we have $\frac{\Delta T}{T}=10^{-5}$. From formula (27)
at $k=1,\quad B_2=0$ we obtain the average energy $\bar E=3\Theta=1.61\cdot
10^{14}$ GeV being of the order of the Grand Unification energy which is not
known exactly. Its estimates vary, say, from $1.9\cdot 10^{14}$ GeV\quad\cite
{K} to $7.5\cdot 10^{15}$ GeV\quad\cite{VT}. We have chosen the estimate (26)
within this range which leads to the CMB temperature fluctuations comparable
with the observed CMB anisotropy values.

For $n=2.5\cdot 10^4,\quad r_0=2.92\cdot 10^{-26}$ cm, $k=1, B_2=0$ from
formula (28) we have $B_4=1.53\cdot 10^{-10}\ll 1$, hence $B_0\approx 1$ due
to formula (8). Since $B_4\ll \frac{k-B_2}{4B_0}$, we can use formula (22) to
calculate the probability of the birth of a hot Universe, which gives
$D=e^{-2.18\cdot 10^{14}}$.

The model predicts a quantum birth of the GUT-scale hot Universe with the
temperatures about those predicted by reheating models and, as a consequence,
the observed CMB anisotropy and plausible amount of monopoles.

\section{Conclusion}

We have considered a possibility of quantum birth of a hot Universe
avoiding the reheating stage. Open and flat universes can be also created due
to quintessence with a negative energy density. The model based on the
GUT-scale false vacuum naturally explains CMB anisotropy and predicts monopole
existence in terms of the initial quantum spectrum of the Universe in the
pre-de-Sitter domain. Thus quantum cosmology proves to have direct
observational consequences, and GUT acquires evidence in its support.

\end{document}